\documentclass{article}

\usepackage{arxiv}

\usepackage[utf8]{inputenc} % allow utf-8 input
\usepackage[T1]{fontenc}    % use 8-bit T1 fonts
\usepackage{hyperref}       % hyperlinks
\usepackage{url}            % simple URL typesetting
\usepackage{booktabs}       % professional-quality tables
\usepackage{amsfonts}       % blackboard math symbols
\usepackage{nicefrac}       % compact symbols for 1/2, etc.
\usepackage{microtype}      % microtypography
\usepackage{lipsum}		% Can be removed after putting your text content

\usepackage{natbib}
\usepackage{doi}

\usepackage{balance}       % to better equalize the last page
\usepackage{graphics}      % for EPS, load graphicx instead 
\usepackage{float}
\usepackage{caption}
\usepackage{adjustbox}
\usepackage{microtype}        % Improved Tracking and Kerning

\author{Amaç Herdağdelen \\
	Core Data Science, Meta\\
	\texttt{amac@herdagdelen.com} \\
	\And 
	Lada Adamic \\
	\texttt{ladamic@gmail.com} \\
	\And 
	Bogdan State \\
	Scie.nz \\
	\texttt{bogdan@scie.nz} \\
}

\def\plaintitle{Community gifting groups on Facebook}

\def\plainkeywords{online community; Buy Nothing; gifting; gift economies}

\makeatletter
\def\url@leostyle{%
  \@ifundefined{selectfont}{
    \def\UrlFont{\sf}
  }{
    \def\UrlFont{\small\bf\ttfamily}
  }}
\makeatother
\def\pprw{8.5in}
\def\pprh{11in}

\hypersetup{%
  pdftitle={\plaintitle},
  pdfauthor={Amaç Herdağdelen, Lada Adamic, Bogdan State},
  pdfkeywords={\plainkeywords},
  pdfdisplaydoctitle=true, % For Accessibility
  bookmarksnumbered,
  pdfstartview={FitH},
  colorlinks,
  citecolor=black,
  filecolor=black,
  linkcolor=black,
  breaklinks=true,
  hypertexnames=false
}
\title{\plaintitle}

\begin{document}
\maketitle

\begin{abstract}
We use de-identified data from Facebook Groups to study and provide a descriptive analysis of local gift-giving communities, in particular buy nothing (BN) groups. These communities allow people to give items they no longer need, reduce waste, and connect to local community. Millions of people have joined BN groups on Facebook, with an increasing pace through the COVID-19 pandemic. BN groups are more popular in dense and urban US counties with higher educational attainment. Compared to other local groups, BN groups have lower Facebook friendship densities, suggesting they bring together people who are not already connected. The interaction graphs in BN groups form larger strongly connected components, indicative of norms of generalized reciprocity. The interaction patterns in BN groups are similar to other local online gift-giving groups, with names containing terms such as `free stuff" and `pay it forward". This points to an interaction signature for local online gift-giving communities.
\end{abstract}

\maketitle
%\maketitleforreal

% ACM Classfication

% Author Keywords
\keywords{\plainkeywords}

\section{Introduction}
From Kula rings to barn raisings, gift economies have long brought communities together and continue to adapt to new forms of social organizations. Even in societies with highly evolved financial systems and pervasive markets, gift economies persist as moneyless generalized exchange systems. Recent years have witnessed an emergence of a new type of gift economy that bring people locally in online spaces~\citeauthor{harvey2018online}. Platforms such as Freecycle, Buy Nothing, Couch Surfing, and BookCrossing provide new opportunities for people to connect to others in novel ways. Online manifestation of these communities provide an interesting opportunity to better understand the dynamics of gifting and community building.

In contrast to market-based economies which rely on dyadic and transactional relations, gift economies are based on a \emph{``web of enduring moral and social commitments within a defined community''}. They need to be able to sustain gift giving activity without a guarantee of direct personal return~\citep{bollier2002stubborn}. One of the mechanisms that help gift economies sustain their cohesive nature is the norm of generalized reciprocity. Generalized reciprocity is a special form of reciprocity in which an exchange happens without any expectation of either direct or delayed reciprocation from the receiver. Colloquially, this phenomenon is known as ``pay it forward'' and it implies the gift giver may receive a similar favorable treatment in the future from another member of the community or that the receiver may ``return the favor'' in the future to someone else who is in need.

Generalized reciprocity is both an antecedent and a consequence of the trust and goodwill among the members of a community. The practice of altruistic exchange without expectation of an immediate reciprocation is both evidence of a community's social capital, which Putnam defines as ``connections among individuals – social networks and the norms of reciprocity and trustworthiness that arise from them'', as well as a direct means through which participants strengthen the very fabric of the community~\citep{putnam2000bowling}. 

Ubiquitous though it may be, generalized reciprocity faces unique challenges in contemporary urban environments. Compared to traditional village communities, the size and complexity of urban environments translate into a wide choice of social contacts. However, this broad potential for social tie formation is undercut by greater difficulty in maintaining and deepening ties in the urban environment. Cities are also typically characterized by high levels of geographic mobility, with people being less likely to live close to family and childhood friends. Indeed, the extent to which the growth of cities has undercut traditional forms of social interaction is one of the defining themes of the sociological analysis of the Modern Age and the Industrial Revolution.~\citep{durkheim2014division} 

A paradoxical combination of abundance and scarcity explains the emergence of a new type of decidedly Internet-age form of generalized reciprocity. Consumers in highly economically developed societies enjoy a surfeit of physical goods, the result of growth spurred by the impersonal forces of market exchange. Economic growth increasingly occurs against a backdrop of resource scarcity, however. In this setting, the logic of sustainability -- the judicious use of limited resources -- has taken on growing strength. In particular, growing numbers of consumers worldwide have faced the challenge of what to do with an increasing number of still-functional products they no longer need.

Modern marketplaces such as eBay and Craigslist provide opportunities to sell items. However, in a world of fast fashion brands and just-in-time manufacturing, the economic returns to trading in used goods have declined. Many owners find it difficult to monetize their no longer needed possessions, especially given the difficult task of assigning fair market value to used items. In the United States, those not interested in selling items, may choose to donate to Goodwill and potentially get a tax deduction. While this is a great option for many, Goodwill only accepts some categories of items. The interaction is impersonal, there is no opportunity to explain what the item is and one does not know who, if anyone, will ultimately use the donated item.

The opportunity of gifting often presents itself as an alluring alternative to selling or bulk donation. In the absence of money, the norms of altruism and gratitude underpin exchange between strangers. The inherent rewards of pro-social behavior, as well as the value imperative of sustainability, thus serve as important motivations for gifting.

A key obstacle to gifting used items is finding those who need them. In tightly-knit communities, the problem of matching between the owners of goods and those in need of them can perhaps be handled through existing social ties. Due to the lower density of networks that can facilitate it, the process of social search is more difficult in urban areas. 

Over time, online spaces have sprung up that can facilitate both giving and receiving items as well as helping to build local community in the process. Examples include the Freecycle Network, which since 2005 coordinates thousands of local volunteer-run groups, some initially using the Yahoo! Groups product. The Nextdoor website allows neighbors to share announcements, questions, as well as for sale and free items. The Facebook groups product supports numerous local groups where items are offered for free and taken, such as moms' groups, ``free stuff" groups, as well as thousands of groups that are part of the official Buy Nothing Project. 

A survey among members of a Freecycle Network chapter showed that individuals' motivations in participating in the movement (in decreasing prevalence) were: decluttering,  self-oriented needs and wants, environmental concerns, and other reasons (helping others, feeling a sense of community, etc.)~\citep{nelson2007downshifting}. \citeauthor{aptekar2016gifts} (2016) found, based on participatory observation and a small-scaled survey of Freecyclers, that among these motivations, decluttering and environmental concerns are more salient. Students who undertook a "buy nothing new, share everything" challenge for a month, as part of a small-scale experiental learning project, were motivated to meet new people and enhance social relations, along with economic and waste-reduction considerations \citep{GODELNIK201740}.

Among these different phenomena, buy nothing groups have initially been almost exclusively organized on Facebook. This presented an opportunity to make one specific case study, with an almost-complete view of how the phenomenon manifests itself on an online platform. We analyze participation in these local online communities over time, especially their rapid growth during the COVID pandemic. We compare buy nothing groups against other Facebook groups which we match on size or localization, showing, for example, that they are more likely to enable local interactions between people who are not already Facebook friends. We also correlate participation in buy nothing groups with offline demographic and other variables at the county level. Finally, we look for other groups with a similar interaction signature as buy nothing groups, and find them to be other instances of community gifting groups, for example, ``free stuff" and ``pay it forward'' groups.

\section{Background}

\subsection{Gifting, Gift Economies, and Local Online Gift-giving}
Gift giving has been traditionally studied within a dyadic framework which focuses on the atomized pair of a gift giver and taker. A gift can be an item, a service or a non-tangible benefit given to someone without expecting compensation. In this framework, gifts are usually picked by the giver specifically for the receiver and there is an established social relation between the pair.

A gift economy or a gift culture is a system where gift exchanges occur within a community. These systems support reciprocity and help maintain social ties within and across communities through rituals and traditions. Some examples that have been studied with ethnographic and anthropological approaches are the potlatch tradition, the big man phenomenon in Melanesia, and the Kula ring phenomenon \citep{weinberger2012intracommunity}. When there is an expectation of direct or delayed reciprocity it can be said that a relationship between the giver and receiver is created or reinforced via gifting.

\citeauthor{giesler2006consumer} identifies a type of gift economy they call the ``consumer gift system'' as a ``system of social solidarity'' that is based on exchanging gifts in structured relationships~\citep{giesler2006consumer} without an expectation of direct reciprocity. According to this view, consumer gift systems go beyond acting ``simply as an aggregate of dyadic gift transactions''  and have emergent communal functions~\citep{weinberger2012intracommunity, corciolani2014gift}. In a way, a gift in a generalized reciprocal setting means creating or reinforcing a relationship between the giver/receiver and the society at large. A study of open source communities~\citep{bacon2012art} found that positive interactions between community members not only build social capital for the two individuals involved, but have ripple effects for the community as a whole.

Take for example, the anonymous, dyadic, but non-reciprocal gifting tradition followed in Mardi Gras celebrations, which \citeauthor{weinberger2012intracommunity} call intracommunity gifting~\citep{weinberger2012intracommunity}. In this tradition, masked members of a parade distribute gifts they bought out of their pocket to watchers. Intracommunity gifting does not maintain or strengthen dyadic relations between specific individuals, but creates bonding in the ``fabric of a community.'' Specifically, it reinforces a sense of belonging to the community and solidarity, allows expressions of gratuity in the moment and generalized forms of gratitude (gratitude that is not felt towards a specific person), and builds status through generosity.

Within this background, we treat local online gift-giving communities as instances of consumer gift systems that function within a local community and enable exchange of resources in the form of gifts. Some examples of these communities are: Freecycle, Buy Nothing, Couch Surfing, and BookCrossing.

 As discussed already, a salient and ubiquitous feature of giftgiving communities is generalized reciprocity, which means that while members are not supposed to offer or give any gifts in return for receiving an item, the expectation is that someone else at some point has and/or will give something to the gifter should they need or ask. This holds for Freecycle \citep{aptekar2016gifts, klug2017gift} and the book exchange network BookCrossing \citep{corciolani2014gift}.
 
 The connection between online gift-giving and generalized reciprocity is not universally accepted in existing literature. Indeed, recent research suggests that some types of pro-social exchange communities may not be conducive to generalized reciprocity as much as it is theorized~\citep{harvey2020prosocial}. As we will discuss next, the norm of generalized reciprocity is explicitly codified and enforced in buy nothing groups, however.

\subsection{Buy Nothing Groups}
 Buy nothing groups on Facebook emerged as part of the Buy Nothing Project\footnote{http://buynothingproject.org/}, founded in 2013 to foster local gift economies and build community ~\citep{clark2020buy}. Since the launch the Buy Nothing project, many local groups have emerged under the self-assigned label of "buy nothing," not all of them formally affiliated with the project. This spontaneous emergence of new community groups is evidence for buy nothing groups having formed a social movement. In an attempt to include all aspects of the phenomenon, our analysis consider all Facebook groups which adopt the ``buy nothing'' label.
 
 Buy nothing groups are structured as local communities each with a defined geographic boundary. As a general rule an individual is expected to be a member of a single buy nothing group whose boundaries include their place of residence. The groups are administered by thousands of volunteer admins, some of whom are supported by the project, who enforce rules and bolster the culture of the movement by posting tips and norms.

Posts within buy nothing Facebook groups are limited to three primary types: offers, requests, and expressions of gratitude. Offers may include things, or one's own time or know-how. Posters are encouraged to let their offers ``simmer'' for a day or so, to allow as many people as possible to express interest. The poster can pick a recipient according to any criteria they wish, including randomly, giving priority to recipients they had not gifted to before, or based on compelling need. The dollar value of a gift is never mentioned, as all gifts are considered priceless \citep{clark2020buy}, from a half-eaten cheesecake to a grand piano. The poster writes a comment notifying the person picked, after which they can coordinate pickup or delivery of the gift via private message.

People can also ask for items they need, often prefacing their requests with markers such as ``before I buy...'' or ISO (in search of). Requests can range from a sprig of parsley to not have to buy a whole bunch, to items of furniture. Sometimes admins will initiate a ``wish-list Wednesday'' post to prompt members to list their wishes.

The third type of post is a gratitude post, wherein people give thanks for gifts they have received, or to the community as a whole. It is not uncommon for people to share some story around a service or item they were gifted.

We also note an additional aspect of the dynamics in buy nothing groups that we call ``vicarious gifting.'' This term represents the fact that both the initial posting of needs or gifts and the response to the post is carried out publicly (within the Facebook group) visible by everyone in the community. Coupled with the strong norms encouraging expressions of gratitude, even bystanders in the groups witness the pro-social consequences of gifting in their community.

The moral undertones in buy nothing groups' mission (sharing out of abundance, strengthening the social fabric, explicitly banning market-based exchange patterns) and the hyper-locality of the groups help its members to identify with their local group and take the interactions beyond atomic commodity exchanges.

\subsection{Relation to social capital}
As we discussed earlier, reciprocity and trust are considered both antecedents and outcomes of robust gift economies. 

Although most people's primary motivation for participating in giftgiving communities may be to efficiently redistribute things and services, organizations such as the Freecycle Network and the buy nothing movement explicitly encourage building community. Through the ``Psychological Sense of Community'' framework, \citeauthor{mcmillan1986sense}~(\citeyear{mcmillan1986sense}) provides a systematic inventory of community-strengthening factors, which are notable in the case of buy nothing groups~\citep{mcmillan1986sense}: 

\textbf{Local membership}: In giftgiving communities, membership is constrained by geographic boundaries due to the face to face interactions required for gifting. In the particular case of buy nothing groups, geographic boundaries that determine membership to specific groups are centrally managed to avoid replicating socio-economic or racial segregation~\citep{bn_geography}. 

\textbf{Norms and rituals}: New members are ritually welcomed in batches by group admins. There are norms around communication patterns, enforced by both admins and members. Expressing interest and its acknowledgement are expected to be done in the comment section of posts without prior permission to send private messages. 

\textbf{Sense of influence and belonging}: Members are welcome to post gratitude posts where they honor who contributed, sometimes with a small story of how they used the item they claimed. Admins regularly create "Wish Wednesday" posts where they encourage people to express their needs and wishes.

We note the similarities between the aforementioned intracommunal gifting and buy nothing movement, despite the differences in structures around gifting. Gifting in buy nothing groups is not anonymous (indeed, both parties are known to the community since gifting occurs publicly). Reciprocity is explicitly forbidden (asking for a need and offering another gift in return is forbidden). However, like Mardi Gras gifting, the giver does not specifically target a gift taker. The gift is not chosen with a particular person in mind and there doesn't have to be an already established relationship between the giver and taker. The ritual of gifting and taking happens in front of everyone to see and potentially develop a sense of generalized, undirected gratitude towards the community. 

Buy nothing groups allow neighborhoods to tap into their community social capital. Unlike the reservoir of social capital people have in their personal social networks, where they can turn to friends to try to offer items or ask for favors, in buy nothing groups individuals can in principle tap into their entire local community. 

In this sense, the gift giving realized through buy nothing groups leads to emergent phenomena. It is more than the sum of the individual social capital of constituent members and their ties. Typically, the problems that require community social capital to solve, are of a type calling for collective action, e.g. pressuring local government to fix infrastructure problems, coordinating participation in school board or other public meetings, or organizing community events. In the case of buy nothing groups, the collective problem being solved is one of inefficiency: underutilized resources piling up or remaining with some individuals, when others in the community could make use of them. Anyone who has ever tried offering a no-longer-used garden or kitchen appliance to friends, or hesitated to bother a friend for a trivial item the friend is not likely to have, will know the limitation of using one-on-one interactions for sharing specific items and favors. This results in an inefficiency that is not solvable by traditional, transaction-based market dynamics. People either hold on to items they no longer need or wouldn't mind lending out, or throw away said items, creating unnecessary waste. 

\section{Data and Methods}
\subsection{Inclusion criteria and reference sample}
To better understand the dynamics of buy nothing groups on the Facebook platform, we selected for analysis those Facebook groups that meet the following criteria:

\begin{itemize}
\item have at least 90\% of the members from the US,
\item have at least 32 members,
\item are either public groups, or are private groups with public visibility (hidden/secret groups were excluded from the analysis). 
\item have at least one comment-based interaction between two US-based members during the 28 days preceding our analysis date (2022-02-25).
\end{itemize}

Starting from this set of groups, we applied additional filters to reach our target and reference sets as follows: 

We identified Facebook groups mentioning ``buy nothing'' or ``buynothing'' in their group name or including a hyperlink to the Buy Nothing Project's web page, http://buynothingproject.org. This resulted in 6254 groups labeled as buy nothing. Buy nothing groups are overwhelmingly private and visible (94\%), and 95\% of them have a membership count between 81 and 3673 (p2.5 and p97.5 of all buy nothing groups).  In order to obtain a comparable reference set of other groups we removed all public groups (in addition to the already filtered out private and hidden groups) and groups with fewer than 81 or more than 3673 members. Finally, we randomly sampled one million groups from the reference set and use it as the baseline for the rest of the analyses. In the subsequent analyses, all group-level features were calculated using de-identified data.

\subsection{Group Characteristics}

Our unit of analysis is a Facebook group. We compute several group-level statistics based on members and member-to-member interaction graphs of the groups.

Group characteristics based on member attributes are distributional summaries of locality and demographic attributes of members and Facebook friendship ties observed among the members of the group.

\begin{itemize}
\item Group size: Number of members in the group ($num\_activated\_members$).
\item Locality: Probability that two randomly picked members are located in the same county ($p\_same\_county$).
\item Gender homophily: Probability that two randomly picked members will be of the same gender, or the gender homophily Blau index ($gender\_blau$).
\item Language homophily: Probability that two randomly picked members will have the same Facebook language (i.e. locale, e.g. en-US for American English) setting, or the language homophily Blau index ($language\_blau$).
\item Age Diversity: Inter-quartile range of the member age distribution (i.e., age difference between the first and third quartiles) ($iqr\_age$).
\item Friendship tie density: Probability that two randomly picked members have a friendship connection on Facebook ($tie\_density$). 
\end{itemize}

Group characteristics based on the member-to-member comment interaction graph, as the name implies, provide a network-level summary of the group based on the directed graph of interactions between members. In this graph, each member who has taken an action (like, comment, post, etc.) in the group in the last 28 days is represented by a node. A directed edge from member A to member B indicates A has replied to a comment or post of B via a comment.

\begin{itemize}
\item Ratio of reciprocal pairs: Ratio of member pairs who have a reciprocal link in the interaction graph to the number of all possible member pairs ($ratio\_reciprocal\_pairs$).
\item Degree inequalities: The Gini coefficient of the in-degree distribution of members (number of other members that wrote a comment to them), and out-degree distribution (number of other members that a member wrote a reply to).
\item Relative size of the largest strongly connected component (SCC): The directed graph allows us to define and compute the strongly connected components in which any member is reachable from another member in the same component, following the links. Ratio of the number of members in the largest SCC to all active members in the graph.
\end{itemize}

The final set of groups and descriptive statistics for the measures we use in the following analyses are given in Table~\ref{tbl:descriptive1}. We will discuss the differentiating aspects of buy nothing groups in the following sections.

\begin{table}[bth]

\begin{tabular}{lll}

\toprule
\textbf{Characteristic} & \textbf{Reference}, N = 1,000,000\textsuperscript{1} & \textbf{Buy Nothing}, N = 5,622\textsuperscript{1} \\ 
\midrule
Number of members & 267 (86, 2,883) & 653 (120, 2,653) \\ 
p(same county) & 0.22 (0.00, 0.87) & 0.82 (0.38, 0.95) \\ 
Gender homophily (Blau) & 0.28 (0.00, 0.50) & 0.25 (0.11, 0.41) \\ 
Language homophily (Blau) & 0.04 (0.00, 0.14) & 0.06 (0.02, 0.16) \\ 
Age diversity (IQR) & 15 (1, 25) & 17 (10, 23) \\ 
Friendship density & 0.04 (0.00, 0.30) & 0.01 (0.00, 0.07) \\ 
Reciprocal pair ratio & 0.005 (0.000, 0.100) & 0.009 (0.002, 0.058) \\ 
In-degree coefficient inequality (Gini) & 0.87 (0.57, 0.98) & 0.69 (0.51, 0.82) \\ 
Out-degree coefficient inequality (Gini) & 0.69 (0.31, 0.91) & 0.59 (0.40, 0.69) \\ 
Largest strongly-connected component ratio & 0.15 (0.02, 0.64) & 0.57 (0.09, 0.77) \\ 
\bottomrule
\textsuperscript{1}n (\%) Median (p2.5, p97.5)\\
\end{tabular}
\caption{Summary statistics of the group-level characteristics in the reference set ($N=1,000,000$) and buy nothing groups ($N=5622$)}
\label{tbl:descriptive1}
\end{table}

\section{Results}
\subsection{Historical Activity and the COVID-19 Pandemic}
\begin{figure}[htp]
\centering
  \includegraphics[width=0.6\columnwidth]{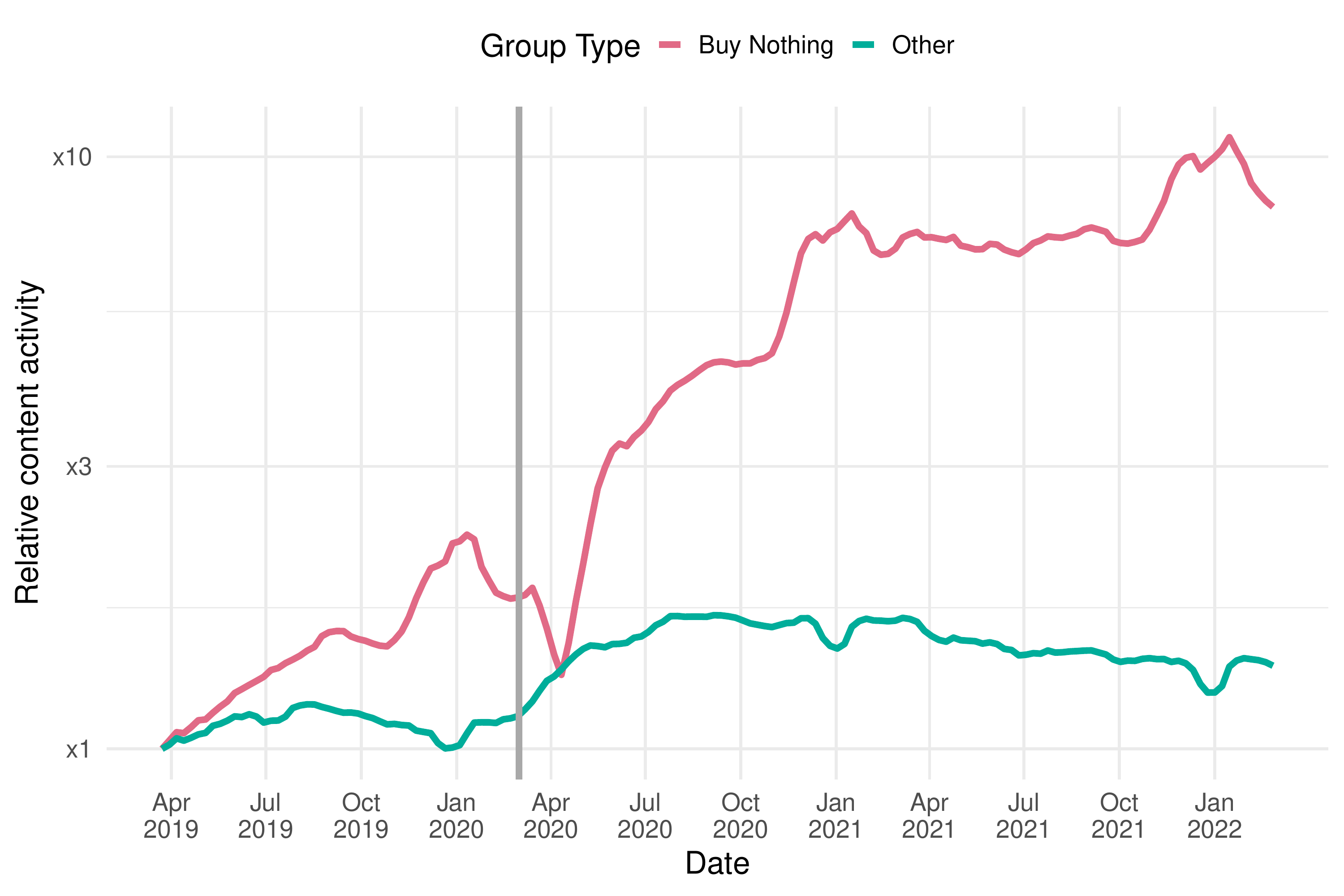}
  \caption{Total number of comments and posts combined in buy nothing groups and the reference set of groups. See text for inclusion criteria. Raw values are omitted due to the confidential nature of the data. Relative values indexed to the values on 2019-04-02, the starting date of the time period. Note the logarithmic scale for the y-axis.}
 ~\label{fig:rel_growth}
\end{figure}

We operationalize the activity and growth by computing the total number of posts and comments generated in groups during a 28-day window. In Fig.~\ref{fig:rel_growth}, we plot the overall activity levels in buy nothing groups and the reference set. Due to the confidential nature of the data, we index the values to 1 for the beginning of the analysis period (2019-04-02), and report the relative growth compared to the levels observed on that date on a semi-logarithmic plot. 

While Facebook groups experienced increases in activity during the COVID-19 pandemic, the growth of buy nothing groups preceded the onset of the COVID-19 pandemic and then accelerated during the pandemic. An initial dip around the time that many parts of the United States initially went into lockdown likely corresponds to early fears that the disease could be transmitted via surfaces. However, activity quickly recovered, possibly as people were spending more time at home and had the opportunity or motivation to improve their home environment, which could include giving away unneeded things, and acquiring new things one needed. Two years into the COVID-19 pandemic, while we see the surge in reference groups is tapering, buy nothing groups continue to increase their activity levels.

\subsection{Geographical Patterns}
\begin{figure}[hbtp]
  \centering
  \includegraphics[width=0.5\columnwidth]{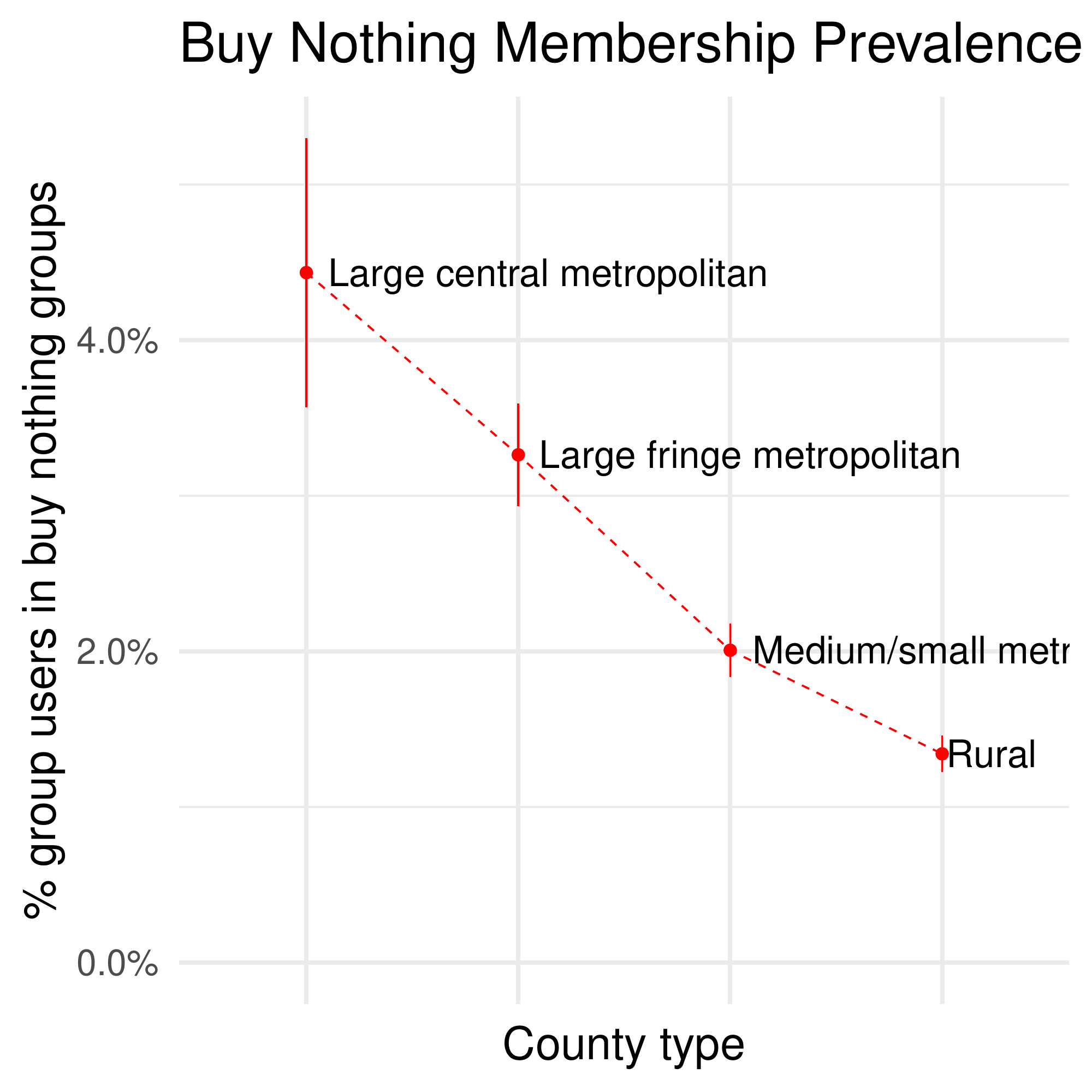}
  \caption{The percentage of users living in a county who are in at least one buy nothing group, averaged over all counties of the given type.}~\label{fig:county_types}
\end{figure}

We define the prevalence of buy nothing membership as the ratio of buy nothing group members living in a county to the number of all Facebook Groups users (people who are members of at least one group) in the same county. In Fig.~\ref{fig:county_types}, we see that prevalence of buy nothing membership shows a clear gradient between urban and rural areas, with the highest prevalence observed in counties designated as urban by the US Census Bureau, and the lowest prevalence observed in rural counties.

\begin{figure*}[htbp]
  \centering
  \includegraphics[width=1.0\linewidth]{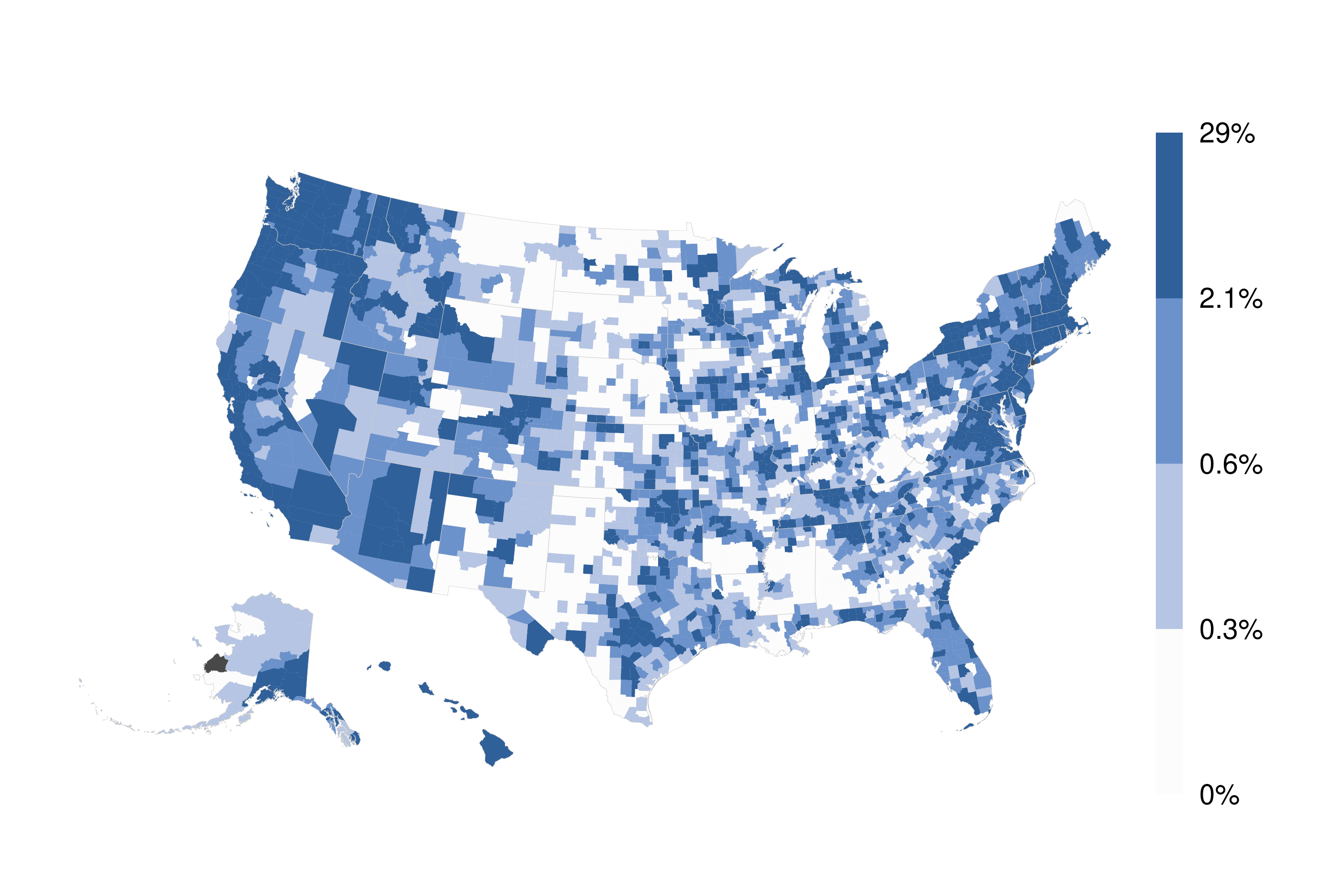}
  \caption{US Counties, grouped by quartiles of buy nothing group membership prevalence. Counties in dark blue are in the top quartile with the highest prevalence, ranging from 2.1\% to 29\% of all Groups users in a county being in a buy nothing group.}~\label{fig:choropleth}
\end{figure*}

The choropleth in Fig.~\ref{fig:choropleth}, reveals that buy nothing membership is mainly an urban and coastal phenomenon in the US.

\subsection{Group Characteristics}

\begin{figure}[]
\centering
  \includegraphics[width=.8\textwidth]{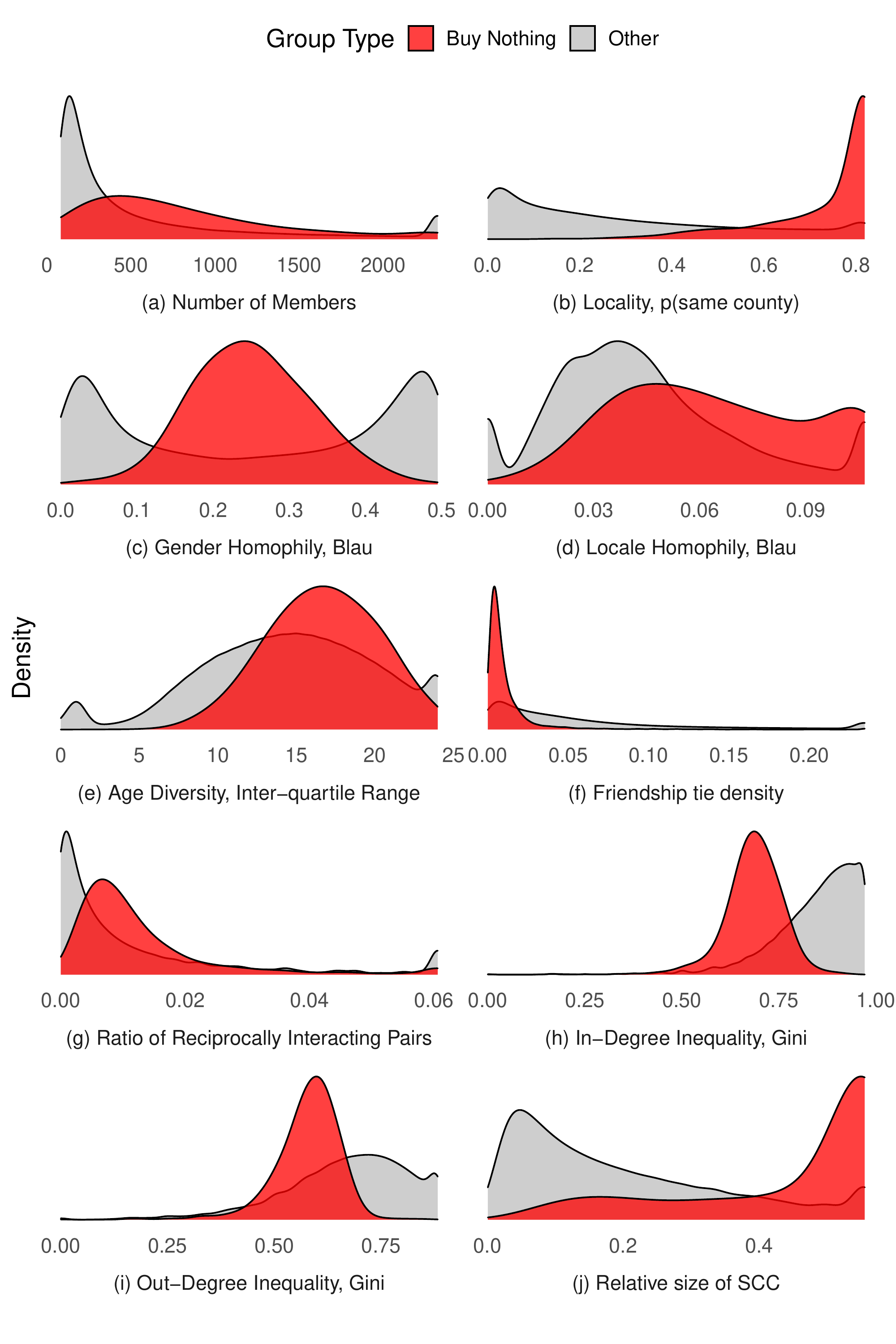}
  \caption{Density distributions of various group characteristics, contrasting the reference group set to buy nothing groups. See main text for details.}
 ~\label{fig:densities}
\end{figure}

In this section, we provide distributional comparisons of buy nothing groups and the reference set. Marginal distributions of the key characteristics described previously are given in Fig.~\ref{fig:densities}. Due to the long-tailed nature of some of the variables, we winsorized each variable by trimming the right-tail at the $95^{th}$ percentile.

\textbf{Size} Even though we matched the range of both the reference set and buy nothing groups to the 2.5th and 97.5th percentile values of the latter, we still observe substantial differences in the group size distributions~(Fig.~\ref{fig:densities}f). Groups in the reference set tend to be smaller than buy nothing groups with much lower mode and average. Since characteristics such as friendship tie density and reciprocal pair ratios are inversely scaled with the number of all pairs, it is important to control for the size to properly characterize the buy nothing groups. We will address this issue in the later sections by employing OLS and controlling for the member count (and its square) and locality score.

\textbf{Locality} The median locality score across the buy nothing groups is 0.82, meaning in half of the buy nothing groups, the probability that two randomly picked members will be living in the same county is over 0.82. In Fig.\ref{fig:densities}b, we see that the distribution of locality scores of buy nothing groups is substantially skewed towards more local compared to the reference set. This is an expected result and consistent with the idea of buy nothing groups creating local gift economies.

\textbf{Language, gender, and age homophily} We observe that buy nothing groups have higher levels of language diversity as evidenced by their lower Blau scores in Fig.~\ref{fig:densities}d (probability that two randomly picked group members will share the same language setting in their app). Similarly, the interquartile range distribution of member ages shows slightly higher values for buy nothing groups (Fig.~\ref{fig:densities}e). Members are less likely to share the same language or be in close age proximity compared to the reference group. Gender homophily exhibits a more nuanced pattern (Fig.~\ref{fig:densities}b. Reference groups have a bi-modal distribution with modes around 0.5 and 1, indicating a higher density of groups with either perfect gender mixing (where the Blau score is at its minimum, 0.5) or consisting of only males or females (where the Blau score is at its maximum, 1). Buy nothing groups, in contrast, occupy a middle ground and a uni-modal distribution. In addition, the gender distribution of members and admins in buy nothing groups are heavily skewed towards more females. On average in a buy nothing group, 86\% of members and 94\% of admins are female (73\% and 72\% respectively in the reference set).

\textbf{Friendship tie density} Buy nothing groups are differentiated with drastically sparser friendship connections among their members, while the reference set exhibits a higher ratio of friendship ties (Fig.~\ref{fig:densities}f). In other words, buy nothing groups tend to bring people together who are not already connected on Facebook via friendship ties. In addition, we observed that a significant portion of those friendships that occur between members of the same buy nothing Groups ($14\%$) occurred after both users joined the group.

\textbf{Reciprocally interacting pairs} In contrast to the sparser friendship tie density, buy nothing groups exhibit higher densities of reciprocally connected pairs of members in the comment interaction graph. In other words, not only do buy nothing groups bring people together who live in the same area but are not connected to each other, but also facilitate direct, reciprocal interactions among these members.

\textbf{Degree inequalities} Both in-degree and out-degree distributions in buy nothing groups exhibit a more equitable pattern, with the Gini coefficient of both distributions being substantially lower than the reference set (Figs.~\ref{fig:densities}h and \ref{fig:densities}i). This is indicating a more decentralized pattern of interactions where the interaction graph is not dominated by a few people either commenting to or receiving comments from many other members.

\begin{figure}[H]
\centering
  \includegraphics[width=0.9\columnwidth]{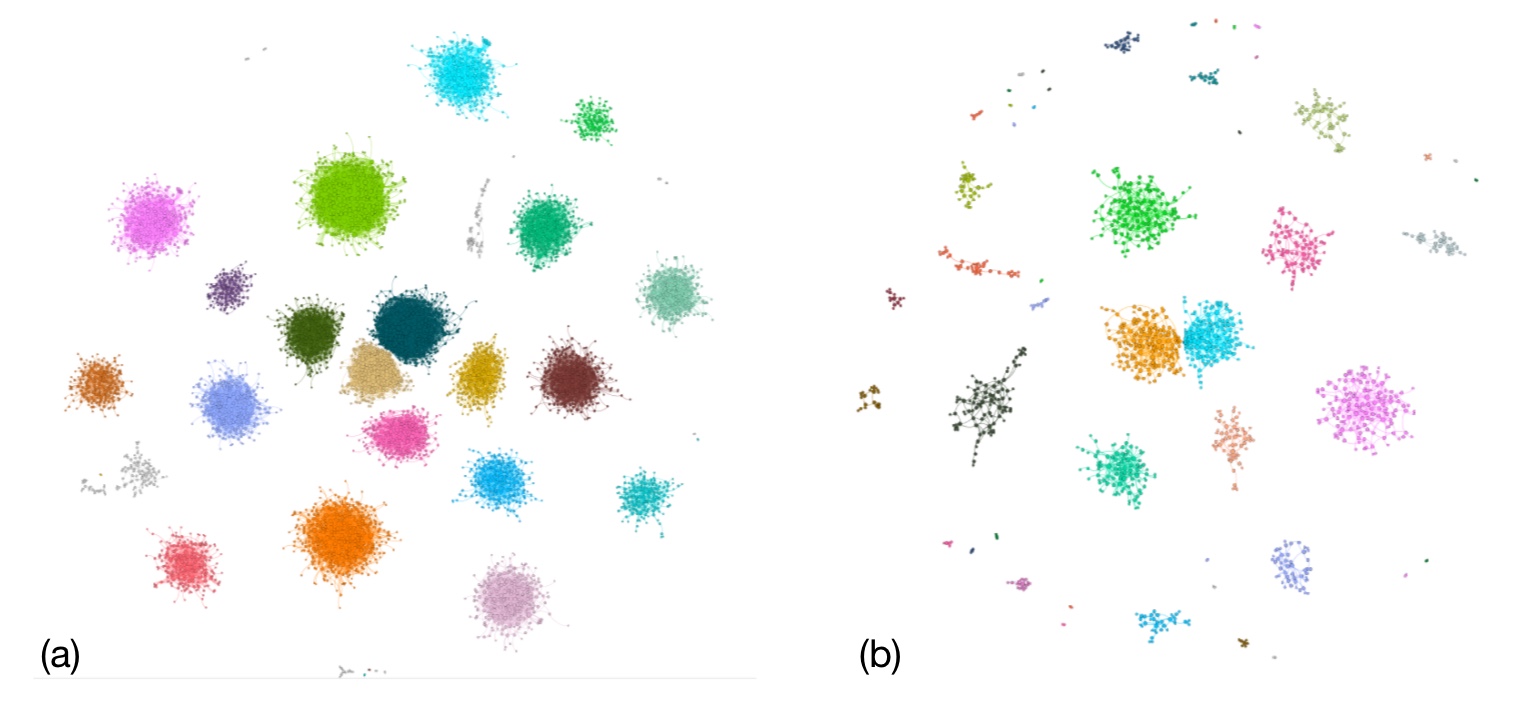}
  \caption{Comment interaction graphs for buy nothing groups vs. other groups matched on size and location, demonstrating higher density and participation in buy nothing groups. Buy nothing groups on the left, matching groups on the right.}
 ~\label{fig:hairballs}
\end{figure}

\textbf{Largest strongly connected component} In tandem with the high reciprocity observed in buy nothing groups, we also see a dominant largest strongly connected component that's, on the average, much larger than what is observed in the reference set~(Fig.~\ref{fig:densities}j). The median relative size of the largest SCC is 0.57 (more than the half), which provides evidence for the norms of generalized reciprocity that is observed in other forms of gifting economies. The difference in density of interaction is also evident in network visualizations of a sample of interaction graphs (see Figure~\ref{fig:hairballs}).

\subsection{Controlling for location, locality, and size}

Figure \ref{fig:densities} provides evidence of important structural differences seen in buy nothing groups compared to other Facebook groups. We find buy nothing groups have higher reciprocal pair ratio, larger largest SCC, and lower friendship density. Given that this figure is only a comparison of marginal distributions, it is important to determine whether these differences are inherent to the structure of buy nothing groups, or whether they are accounted for by covariates unrelated to the giftgiving nature of the groups, such as their size and locality.

Our reference set contains both local and non-local groups and presumably contains family-/friendship-based groups which tend to be smaller in size and denser in friendship ties. In order to tease out the effects of these group covariates, we created a matched reference set to control for the location, locality, and size of the group. We first identified all groups in the buy nothing and reference set with a locality score higher than 0.25 (which corresponds to the theoretical minimum locality score a group can get while having a majority of its members in the same city).  Then we identified all cities with at least two buy nothing groups and ten non-buy nothing groups (where the city is the majority city among the members of the groups). This procedure gives us a set of groups that can be matched on city and size, while still looking at the differences in the key characteristics we are analyzing in this study.

We use ordinary least squares to estimate the effect of being a buy nothing group on the key group characteristics. The models are specified as 

$$y_i = \beta_0 + \beta_{bn} \delta_{i} + \mathbf{\beta_{i}} \mathbf{X_i} + \epsilon_i$$

where $y_i$ are the outcome variables that correspond to the winsorized group characteristics (tie density, degree inequalities, gender/age/language diversities, relative size of the largest SCC, and ratio of reciprocally connected pairs) for group $i$. $\delta_{i}$ is a dummy variable that takes the value 1 if the group $i$ is a buy nothing group and 0 otherwise. The vector $\mathbf{X_i}$ encodes the control covariates for the group which include a dummy variable for the city the group is associated with, size of the group (and its square), and the locality score of the group. 

\begin{figure}
\centering
  \includegraphics[width=0.9\columnwidth]{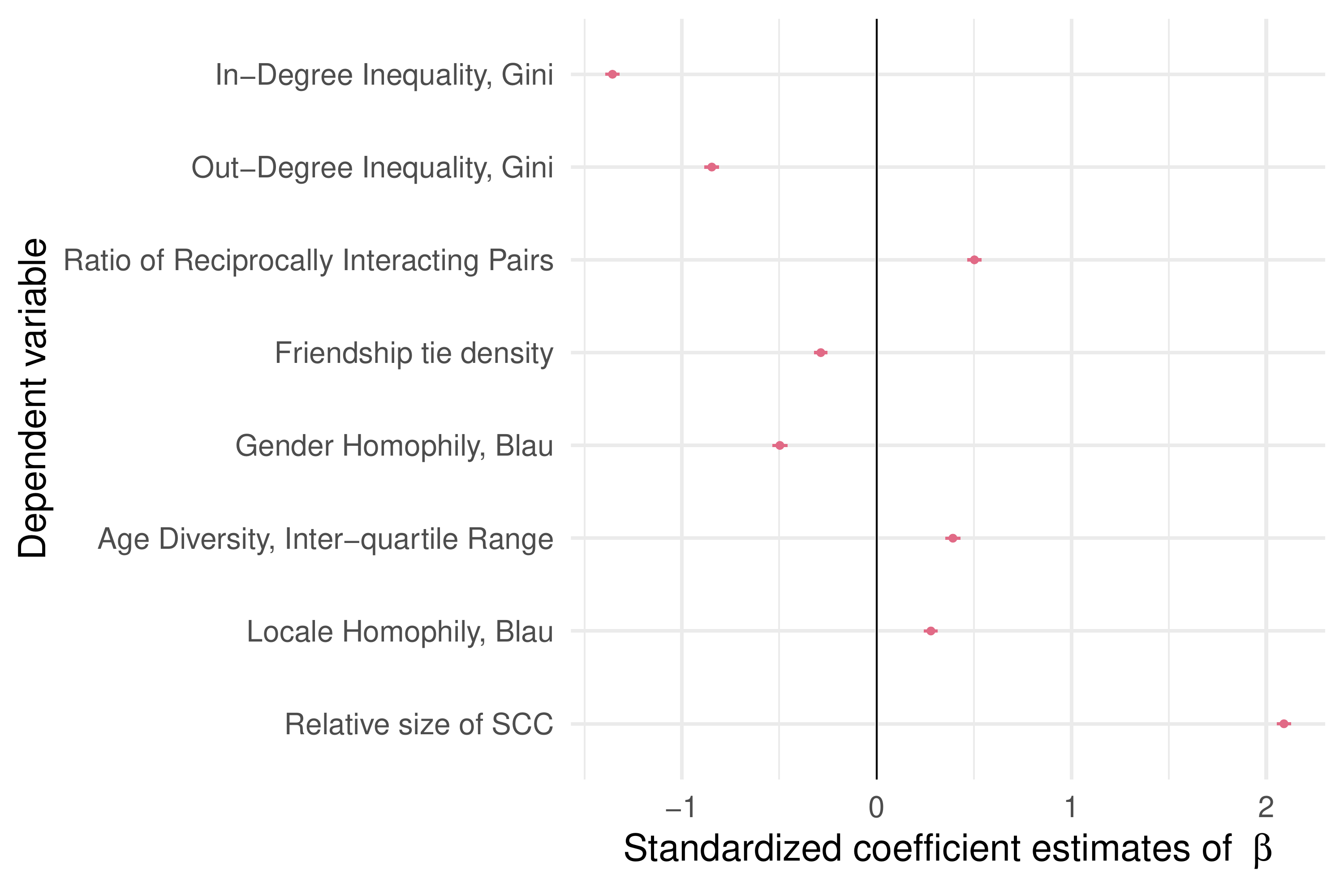}
  \caption{Effect of buy nothing dummy variable on key characteristics in cities with at least 10 reference groups and at least 2 buy nothing Groups. Group controls are city fixed effects, number of members (and its square), $p\_same\_city$, and $p\_same\_county$. Coefficients are estimated for standardized outcome variables (divided by 1 standard deviation).}
 ~\label{fig:coefficients}
\end{figure}

This specification allows us to control for both contextual effects the location may have on groups (by employing city fixed effects), and group-level covariates of size and locality. The estimate of the coefficient $\beta_{bn}$ provides a way to quantify the difference in the outcome variable between buy nothing groups and the other groups. Since the dependent variables are normalized, the coefficient can be interpreted as the difference in standard deviations. Note that, by creating a matched set, we are losing the representativeness of the data and we cannot interpret the coefficient estimates as unbiased estimates for all buy nothing groups. We are using this exercise to provide directional evidence. The estimated coefficients are given in Fig.~\ref{fig:coefficients}.

With the exception of language diversity, all of the estimated coefficients corroborate our visual interpretations in the previous section. After controlling for the majority city, group size, and locality score of the group, buy nothing groups exhibit higher levels of equity in both in- and out-degree distributions, have more reciprocally interacting member pairs, fewer friendship ties among members, provide more gender and age mixing, and have a larger strongly connected component. language homophily, on the other hand seems to be higher in buy nothing groups according to this model.

\subsection{Correlations with offline social capital measures}

If gifting groups are a way for communities to tap into their social capital, one might expect areas of the United States with high measures of social capital to also be host to a disproportionate amount of buy nothing activity. On the other hand, lack of existing norms of reciprocity and social capital in an area may foster wider adoption of gifting groups as a substitute for the missing social capital.

In order to study the geographic correlation between buy nothing group prevalance and social capital, we use two county-level social capital indicators estimated by the Social Capital Project of the Senate’s Joint Economic Committee \citeauthor{JEC2018}. These are \emph{community health} and \emph{institutional health}.

Community health includes per capita rates of volunteering, attending public meetings and protests, and working with one's neighbors to improve community. Institutional health includes voting rates in presidential elections, mailing in census questionnaires, and the share of adults with at least some degree of confidence in corporations, the media, and public schools. 

Both indices were uncorrelated with the percent of group users in a county who were members of a buy nothing group.

There are, however, environmental factors that might influence the growth of buy nothing groups in an area. We found, for example, that buy nothing groups are more prevalent in urban counties ($\rho = 0.32$), with a higher population ($\rho=0.43$), and population density ($\rho=0.37$). This may be simply because higher population density enables critical mass. It could also be the case that in rural areas neighbors already know one another and would have less need to gift items through buy nothing groups. 

\begin{figure}[h]
  \centering
  \includegraphics[width=0.5\columnwidth]{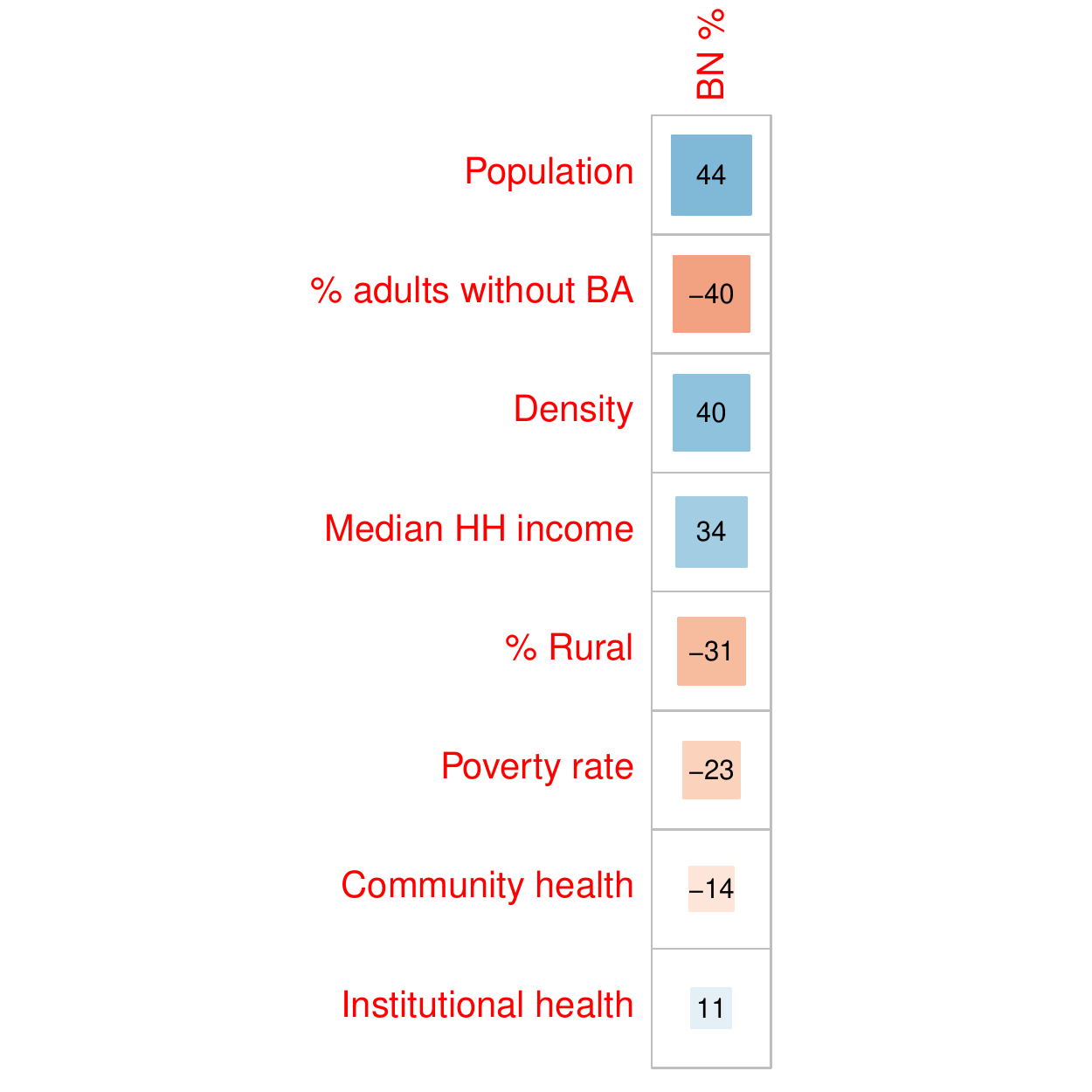}
  \caption{Comparison of offline social capital estimates and activity in buy nothing groups by county in the United States.}~\label{fig:heatmap}
\end{figure}

 Buy nothing membership prevalence is more represented in higher income counties ($\rho = 0.36$), with a lower poverty rate ($\rho = -0.33$), and lower percentage of adults without a college degree ($\rho = -0.42$).  
 
It would be interesting to follow up in future work to examine why buy nothing activity correlates with median household income at the county level. One explanation may be that excess consumption correlates with income, which would mean that individuals have a greater number of items to ``dispose" of. Another may be that at lower income, individuals may choose to sell rather than gift items to supplement their income.

\subsection{Other groups similar to buy nothing groups}
\begin{figure}
  \centering
  \includegraphics[width=0.5\columnwidth]{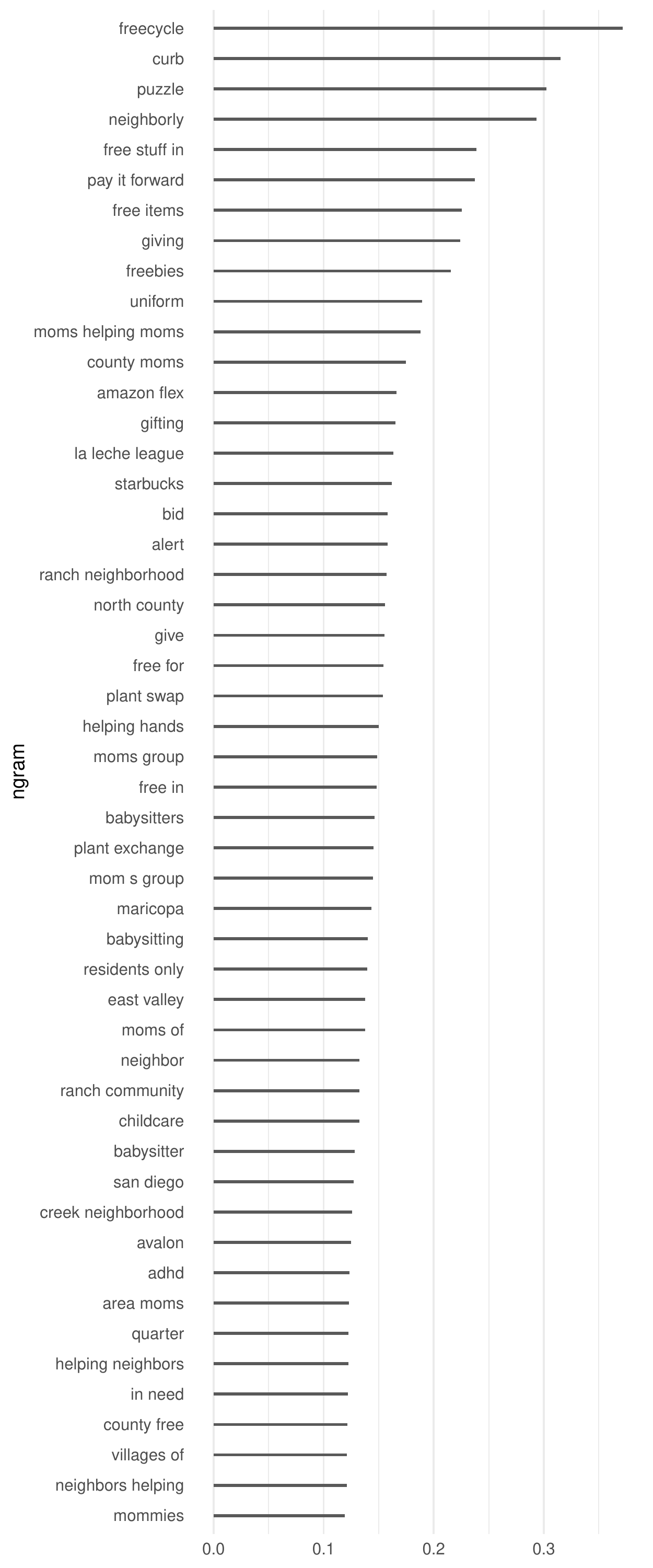}
  \caption{Highest-scoring 50 ngrams for distinguishing between BN groups and a randomly-sampled reference set}~\label{fig:ngrams}
\end{figure}

In order to assess whether the differentiating characteristics we analyzed in this study are due to idiosyncratic mechanisms and norms associated with the buy nothing movement or are indeed associated with the more general notion of local gift economies, we identified other non buy nothing groups that are similar to the buy nothing groups in our sample across these characteristics. In order to do that, we trained a Gradient Boosting Machine (xgboost, see \citeauthor{chen2015xgboost}) using structural and interactional patterns to predict if a group is buy nothing or not (notably we do not employ any content-based features such as posts or name of the groups). We applied the model to the reference set and evaluated a classifier prediction score for each group. We extracted ngrams from the names of the groups, and for each ngram computed the average prediction score of the groups that contain the ngram in their names. Fig.~\ref{fig:ngrams} lists the highest-scoring 50 ngrams (after removing lower-order ngrams that are included in longer ngrams).

It is striking that groups that mention ``Freecycle'', ``Curb (alert)'', ``pay it forward``, ``moms helping moms'', ``freebies'', ``plant swap'', etc. receive high prediction scores by the classifier. These results suggest that local solidarity and free sharing groups do indeed share similar structural and interactional patterns with the more specific class of buy nothing groups, increasing our confidence in the generalizability of our results towards other gifting groups.

\section{Limitations}
Some limitations that hamper the generalizability of our results are as follows:

Buy nothing groups are only one instantiation of local online gift communities, many of which have access to a central support system for the volunteer admins. In the previous section, we provided evidence showing that some other types of local solidarity and sharing groups exhibit structural and interactional similarities to buy nothing groups. However, identification of all sharing and local solidarity groups was beyond the scope of our study and we cannot generalize our results to all other forms of sharing communities.

Furthermore, we are focusing only on open and closed groups, excluding secret groups. Our analysis may be missing more tightly-knit communities that prefer to organize in secret groups that are invitation-only and not visible to members of the public.

The observation period coincided with a global pandemic of COVID-19, and studying the effects of lockdowns, avoidance of social gatherings and other impacts of the pandemic was out of scope. The extent of how the pandemic shaped our observations is an open question.

Our reference group set was picked to match the size and locality attributes of buy nothing groups very coarsely and we haven't made attempts to come up with more specific baseline groups. In the future, more specific research questions about the nature of local online gifting communities may warrant more specific and carefully designed matching reference sets.

An important limitation about the interactional patterns is that we do not have direct signal on the direction of gifting and whether commenting and posting activities always result in gifting behavior. For example, we haven't quantified the prevalence of asking posts versus offering posts. In addition, the comment interaction graph we analyze to support the generalized reciprocity is only indicative as a proxy because we do not have the actual graph of gifts flowing among the members.4

\section{Discussion and Future Work}

In this paper, we provided a descriptive summary of compositional and interactional patterns observed in buy nothing groups on Facebook, supporting and quantifying several theoretical considerations of local online gifting communities.  

Comment and post volume in buy nothing groups has grown almost five-fold during the course of the COVID-19 pandemic. 

The interaction patterns in these groups based on member to member comment graphs are dense, and exhibit more equal participation than in other local groups of similar size. The largest strongly connected component is larger in buy nothing groups, with a median of 57\% of total group size. 

Many groups are founded on the same foci in life that friendship stems from, through continued exposure: sharing the same school, place of worship, hobby or workplace. It is therefore unsurprising that many Facebook Groups have relatively high friendship density: meaning that many members are already Facebook friends. In contrast, buy nothing members do not necessarily share a common context, besides living in the same geographical area, and consistently the friendship density in buy nothing groups is lower, as shown in panel f of Figure~\ref{fig:densities}.

Buy nothing groups may very well be part of reconnecting neighborhoods: they allow people to tap into the social capital of their community, without needing to form and maintain individual friendship ties, although those can certainly occur as a result of buy nothing interactions. Approximately 14\% of friendship ties among members of the same buy nothing groups were formed after both users joined the group.

Interestingly, the rate of participation in buy nothing groups does not correlate with offline measures of social capital. Prevelance of membership in buy nothing groups is higher in urban, highly-educated, high-income areas. Yet, the buy nothing movement is clearly tapping into the need of individuals not just to get rid of and acquire stuff, but also to interact with their community. In doing so, the buy nothing movement is likely also creating new social capital, through the growth of trust in one's neighbors, the new connections formed, and the many lending libraries and item ``trains'' that are now community resources. Causal effects of buy nothing group activity on community health and social capital would be an interesting topic of further study.

Finally, when we look at other Facebook groups that are similar to buy nothing groups along the dimensions of interaction graph features and other group characteristics, we find groups that contain the terms indicating other pay-it-forward style local gifting communities. These keywords include, but are not limited to: ``freecycle'', ``pay it forward'', ``moms helping moms'', ``plant swap'', and ``neighbors helping neighbors''. These findings suggest that the characteristics we describe in this paper are not only salient attributes of buy nothing groups, but capture some general aspects of local online gifting communities.

% BALANCE COLUMNS
\balance{}

% REFERENCES FORMAT
% References must be the same font size as other body text.
%\bibliographystyle{SIGCHI-Reference-Format}
%\bibliography{references}
%\printbibliography
\bibliographystyle{ACM-Reference-Format}
\bibliography{references}
\end{document}